# Interactions of Atomic and Molecular Hydrogen with a Diamondlike Carbon Surface: $H_2$ Formation and Desorption

Masashi Tsuge, Tetsuya Hama, Yuki Kimura, Akira Kouchi, and Naoki Watanabe

Institute of Low Temperature Science, Hokkaido University, Kita-19, Nishi-8, Kita-ku, Sapporo 060-0819, Japan: tsuge@lowtem.hokudai.ac.jp

Abstract

The interactions of atomic and molecular hydrogen with bare interstellar dust grain surfaces are important for understanding $H_2$ formation at relatively high temperatures (> 20 K). We investigate the diffusion of physisorbed H-atoms and the desorption energetics of $H_2$ molecules on an amorphous diamondlike carbon (DLC) surface. From temperature-programmed desorption experiments with a resonance-enhanced multiphoton ionization (REMPI) method for $H_2$ detection, the $H_2$ coverage-dependent activation energies for $H_2$ desorption are determined. The activation energies decrease with increasing $H_2$ coverage and are centered at 30 meV with a narrow distribution. Using a combination of photostimulated desorption and REMPI methods, the time variations of the surface number density of $H_2$ following atomic and molecular hydrogen depositions are studied. From these measurements, we show that $H_2$ formation on a DLC surface is quite efficient, even at 20 K. A significant kinetic isotope effect for $H_2$ and $D_2$ recombination reactions suggests that H-atom diffusion on a DLC surface is mediated by quantum mechanical tunneling. **In astrophysically relevant conditions, $H_2$ recombination due to physisorbed H-atoms is unlikely to occur at 20 K, suggesting that chemisorbed H-atoms might play a role in $H_2$ formation at relatively high temperatures.**

# 1. Introduction

Molecular hydrogen ($H_2$) is the most abundant molecule in the universe and plays important roles in many astronomical processes. Because the gas-phase formation of $H_2$ is too slow to account for its high abundance in the interstellar medium, it is generally considered that $H_2$ is formed on dust grain surfaces (Gould & Salpeter 1963). The $H_2$ formation on dust grain surfaces proceeds through elementally processes: H-atom adsorption, diffusion, and encountering another H-atom. Since $H_2$ recombination between two H-atoms weakly bound to the surface tends to be a radical-radical barrier-less reaction, the $H_2$ formation rate is generally limited by the diffusion of H-atoms when the surface coverage is small. Thus, the determination of the H-diffusion rate as well as its mechanism at very low temperature (~10 K) is crucial to understanding the $H_2$ formation process on dust grains (Hollenbach & Salpeter 1971). The temperature dependence of $H_2$ formation is another important issue in the astronomical context because the $H_2$ formation mechanism at relatively high temperature (> 20 K) remains unclear.

Elementary processes associated with $H_2$ formation on dust grain surfaces have been studied both experimentally and theoretically (Hornekær 2015; Vidali 2013; Wakelam et al. 2017), especially for the amorphous solid water (ASW) surface (Hama & Watanabe 2013). In cold molecular clouds with temperatures of approximately 10 K, where a number of molecules are produced, dust grains are covered with ASW. One important conclusion from these extensive investigations is that $H_2$ formation on the ASW surface efficiently occurs only for a temperature range up to 16 K (Perets et al. 2005); at higher temperatures, H-atoms desorb from the surface before they encounter each other. $H_2$ formation at a relatively higher temperature region, i.e., ~100 K, cannot be attributed to formation on the ASW surface, as a sufficiently low temperature is required for the dust grains to be covered with ASW (Oba et al. 2009). Carbon monoxide is also important molecule that covers dust grains (Kimura et al. 2018; Pontoppidan et al. 2003; Pontoppidan 2006), but its condensation to bare grain surfaces also requires lower temperatures (< 20 K). Therefore, we should consider other $H_2$ formation mechanisms that work at higher temperatures, which are relevant to $H_2$ formation such as in diffuse clouds.

Reactions on the bare surfaces of dust grains may be responsible for the $H_2$ formation at relatively high temperatures. Bare dust grains mainly consist of silicate or carbonaceous materials, though the specific characteristics remain unclear. Experimental studies about $H_2$ formation on silicate surfaces have been reported (He, Frank, & Vidali 2011; Perets et al. 2007; Pirronello et al. 1997; Vidali & Li 2010; Vidali et al. 2009). Pirronello and coworkers studied $H_2$ (actually HD) formation on silicate surfaces by means of temperature programmed desorption (TPD) and reported the barrier height for H-atom diffusion to be 25 meV for polycrystalline silicate and 35 meV for amorphous silicate (Perets et al. 2007; Pirronello et al. 1997). From a rate equation model, the authors suggested that, under astronomical conditions, efficient $H_2$ formation is expected only in a limited temperature range from 8 to 13 K. Katz et al. used a more simplified model to formulate $H_2$ formation efficiency and reached a similar conclusion (Katz et al. 1999). Cazaux and Tielens further developed a model whereby quantum-tunneling diffusion and chemisorption are taken into account (Cazaux & Tielens 2004). Although quantum-tunneling from a physisorbed site to a chemisorbed site has yet to be verified experimentally, their model suggests that $H_2$ formation is efficient until quite high temperatures (~500 K).

$H_2$ formation on carbonaceous surfaces has also been studied in a manner similar to that on silicate surfaces. For a flat and nonporous graphitic amorphous carbon sample, Pirronello et al. studied HD formation by TPD and determined the activation energy for the overall process of atomic diffusion and molecular desorption to be ~45 meV (Pirronello et al. 1999). Subsequently, their TPD data were used to derive temperature- and H-atom-flux-dependent $H_2$ formation efficiencies; at a flux relevant for astronomical conditions, Katz et al. suggested that $H_2$ formation is efficient for the temperature range 10–14 K (Katz et al. 1999). According to Monte Carlo simulations performed by Cuppen and Herbst, the temperature range for high-efficiency $H_2$ formation increases, to 30 K, for higher surface roughness (Cuppen & Herbst 2005). The $H_2$ formation between chemisorbed and gas-phase H-atoms via the Eley–Rideal abstraction mechanism was reported for a hydrogenated porous, defective, aliphatic carbon surface (Mennella 2008), and such a reaction should be efficient for grain temperatures up to 300 K. **Martín-Doménech et al. performed vacuum ultraviolet photolysis of hydrogenated amorphous carbon and suggested that photo-produced $H_2$ in the bulk of the dust particles can diffuse out to the gas phase (Martín-Doménech, Dartois, & Muñoz Caro 2016)**.

Another important class of amorphous carbon is tetrahedrally bonded amorphous carbon, ta-C, in which $sp^3$ bonds dominate (**Ong et al. 1995**; Robertson 2002; Schultrich 2018; **Voevodin & Donley 1996**). This class of carbonaceous material is often denoted as diamondlike carbon (DLC). The existence of DLC (or nanodiamond) in interstellar media has been inferred from the observation of a widespread broadband emission in the 650–800 nm range, which is called the extended red emission (ERE) (Duley & Williams 1988; Lewis, Anders, & Draine 1989). Among the proposed carriers of the ERE including hydrogenated amorphous carbon (Duley & Williams 1988; Witt & Schild 1988), quenched carbonaceous composites (Sakata et al. 1992), $C_{60}$ (Webster 1993), and silicon nanoparticles (Ledoux et al. 1998; Witt, Gordon, & Furton 1998), it has been demonstrated that nanodiamonds (size ~100 nm) containing nitrogen-vacancy defects show a strong resemblance to the astronomical spectra of ERE (Chang, Chen, & Kwok 2006). Regardless of possible ubiquitous nature of DLC in the interstellar media (Allamandola et al. 1993), the interactions of atomic and molecular hydrogen with the DLC surface have yet to be studied experimentally.

In this work, we prepared a hydrogen-free DLC sample by means of pulsed laser ablation. For this sample, we investigated the $H_2$ binding energy (≈ activation energy for $H_2$ desorption) and $H_2$ formation processes at low temperatures. From the TPD measurements, the coverage-dependent activation energies for $H_2$ desorption are determined. By measuring the surface number density of $H_2$ *in situ*, we will show that $H_2$ formation on a DLC surface is efficient even at 20 K, and based on a significant isotope effect, we will discuss the H-atom diffusion mechanism (i.e., thermal hopping and/or quantum-tunneling).

## 2. Experiment

### 2.1. Experimental setup and procedure

The experiments were performed by using a specially designed apparatus named Reaction Apparatus for Surface Characteristic Analysis at Low-temperature (RASCAL) at the Institute of Low Temperature Science, Hokkaido University (Hama et al. 2012; Watanabe et al. 2010). RASCAL consists of a main chamber, a linear type time-of-flight (TOF) mass spectrometer, a doubly differentially pumped molecular and atomic hydrogen source, as well as two laser systems for the photostimulated desorption of surface adsorbates and for the ionization of desorbed $H_2$ molecules.

The main chamber was evacuated to ultra-high vacuum conditions (~ $10^{-8}$ Pa) using two turbo molecular pumps. A mirror-polished Al substrate (30 mm diameter), which was connected to the cold head of a closed-cycle He refrigerator, was located at the center of the chamber and can be cooled to ~5 K. The single beam line for both molecular and atomic hydrogen was composed of two differentially pumped stages. Collimators with 1–2 mm apertures were located between the first and second chambers and between the second and main chambers. $H_2$ gas was introduced to the first chamber through a Pyrex tube and then cooled to 100 K in an Al pipe cooled by another He refrigerator. An atomic hydrogen beam was produced in a microwave-discharge plasma in the Pyrex tube, and the dissociation fraction was approximately 70%**; see Hama et al. 2012 for the method to determine dissociation fraction**. The typical pressures of the first, second, and main chambers during the operation were $1 \times 10^{-3}$, $1 \times 10^{-5}$, and $5 \times 10^{-8}$ Pa, respectively.

A DLC sample was prepared in the main chamber by pulsed laser ablation (**Ong et al. 1995**; Schultrich 2018; Voevodin & Donley 1996). During preparation, a carbon target (99.99% graphite, Nilaco Corp.) was located near the Al substrate, at approximately 40 mm in distance and a 45° angle. The rotating graphite target was irradiated with a pulsed laser beam (532 nm wavelength, 8 ns pulse width, 10 Hz repetition rate, and 20 mJ pulse energy) focused with an $f$ = 300 mm lens. Assuming a laser spot size of 1 mm, the laser intensity was estimated to be $2.5 \times 10^8$ W cm$^{-2}$. Typically, laser ablation was performed for 6000–9000 pulses, resulting in an approximately 20–30 nm thick sample on an Al substrate.

Hydrogen molecules on the Al substrate or DLC sample were detected by a combination of photostimulated desorption (PSD) and resonance-enhanced multiphoton ionization (REMPI) methods, PSD-REMPI; the technique as analogous to the procedure used in former experiments, and the details were previously described (Hama et al. 2012; Kimura et al. 2018; Watanabe et al. 2010). The experimental procedures are schematically shown in Figure 1. In brief, a small fraction of $H_2$ was sequentially sublimated by weak radiation from a PSD YAG laser (532 nm, 10 Hz, and < 10 µJ pulse$^{-1}$), rotational-state selectively ionized by (2 + 1) REMPI via the $E,F^1$ ($v' = 0, J' = J''$) ← $X^1$ ($v'' = 0, J'' = 0$ or 1) transition (Huo, Rinnen, & Zare 1991; Pomerantz et al. 2004; Rinnen et al. 1991) at approximately 1.0 mm above the sample surface, and detected by the TOF mass spectrometer. Laser radiation in a wavelength range of 201–203 nm with a pulse energy of 100–200 µJ was provided from a dye laser pumped with a Nd$^{3+}$:YAG laser, with subsequent frequency doubling and mixing in potassium dihydrogen phosphate and beta barium borate crystals. The delay time between PSD and REMPI laser irradiations was set to coincide with the moment when the $H_2$ signal reaches the maximum value.

For $H_2$ molecules on the ASW and Al substrate, the photostimulated desorption is thought to occur due to phonons excited by 532 nm PSD laser radiation (Fukutani et al. 2005; Hama et al.

2012). One of the characteristics of phonon-mediated PSD is that the signal intensity is proportional to the intensity of the PSD laser. However, in the case of PSD from the DLC sample, the $H_2$ signal intensity was approximately proportional to the square of the PSD laser intensity (up to 200 μJ pulse$^{-1}$), indicating a different mechanism for desorption. Moreover, the desorption efficiency did not correlate with the thickness of the DLC sample; for example, a similar desorption efficiency was observed for ~20 and ~100 nm thick samples. This phenomenon is a strikingly different from the desorption of $H_2$ on the ASW that was prepared on an Al substrate, where the $H_2$ signal intensity decreases as a function of ASW thickness (Hama et al. 2012). We believe the desorption of $H_2$ from a DLC sample is related to the absorption of 532 nm radiation by the DLC sample (Malshe et al. 1990; Robertson 2002). However, we were unable to distinguish photochemical and photothermal processes, and the topic is beyond the scope of this paper.

Temperature-programmed desorption (TPD) spectra were rotational-state-selectively recorded by detecting desorbing $H_2$ by the REMPI method. In the experiment, molecular or atomic hydrogen was deposited onto the DLC sample at 8 K for a certain period, and the sample was warmed with a typical ramp rate of 4 K min$^{-1}$. In this paper, we call the spectra obtained by this method REMPI-TPD spectra.

**2.2. Characterization of samples**

For the characterization of samples by a transmission electron microscope (TEM), a thin DLC film was prepared on a KBr pellet by the same pulsed laser ablation method and was then exfoliated on a water surface by a combination of the surface tension of water and the dissolution of the KBr substrate. The thin DLC film floating on the water surface was scooped by a standard copper grid for TEM observation. A TEM apparatus (JEOL, JEM-2100F) was used with a field-emission gun at an acceleration voltage of 200 kV. Transmission electron micrographs are shown in Figure 2. The inhomogeneity of the contrast shown in Figure 2a most likely originates from the undulations of the KBr pellet, which was prepared by pressing KBr powder. The electron diffraction pattern presented in Figure 2b clearly shows two Debye rings, corresponding to 2.12 and 1.16 Å, attributed to DLC (Mōri & Namba 1984) and does not show a strong Debye ring originating from the 3.4 Å of graphite (Bacon 1951).

The H-atom signal due to chemisorbed H-atoms was undetectable for a fresh DLC sample, indicating that we successfully prepared a hydrogen-free DLC sample. However, we found that after exposure to hydrogen, H-atoms could be detected when the DLC sample was irradiated with a relatively intense PSD laser (> 200 μJ pulse$^{-1}$) for the investigated temperature range of 8–300 K; H-atoms desorbed from the surface can also be detected by the REMPI method (Zumbach et al. 1997). These H-atoms might be due to chemisorbed ones; in other words, our sample could be partially hydrogenated, and the signal intensity increased as a function of the $H_2$ or H-atom fluence applied to the surface. Hydrogenation occurred at all the investigated surface temperatures (8–300 K) and impinging atomic or molecular hydrogen beam temperatures (100–300 K). Thus, our DLC sample aged during the experiments. At 300 K, the chemisorbed H-atom signal started to saturate after applying a $H_2$ fluence of up to $1 \times 10^{16}$ cm$^{-2}$. To estimate the fraction of hydrogenated sites, we tried to follow hydrogenation by infrared spectroscopy. For this purpose, the collimators in the hydrogen source were removed so that whole area of the substrate was irradiated with $H_2$ or H-atoms. However, we were unable to detect any IR absorption in the region of the CH-stretching vibration; this behavior is in contrast to the hydrogenation of a porous, defective, aliphatic carbon surface (Mennella 2008), where the infrared signature of the sp$^3$-hybridized CH bonds was readily observed. These results indicate that the fraction of hydrogenated sites on our DLC sample is quite low ($10^{-5}$ or smaller). The experimental results presented in this paper are unaltered between a newly prepared DLC sample and an aged one. **Thus, the $H_2$ formation via the Eley–Rideal mechanism between chemisorbed and gas-phase H-atoms would be very minor in our experiments.**

## 3. Results

### 3.1. Estimation of H$_2$ flux and adsorption site density

The flux of the H$_2$ beam was estimated according to the method reported by Hama et al. (Hama et al. 2012), in which H$_2$ was continuously deposited onto an Al substrate at 8 K and detected by the PSD-REMPI method. The sum of H$_2$ signals (from $J'' = 0$ and 1 levels) started to saturate after 900 s of deposition (Figure 3a). Because a multiple-layer deposition of H$_2$ does not occur at 8 K, this saturation indicates that the coverage approached unity. Assuming an adsorption probability of ~100% and the site density of Al to be $1.2 \times 10^{15}$ atoms cm$^{-2}$ (Wyckoff 1931), we estimate the flux to be $1.2 \times 10^{15}$ atoms cm$^{-2}$ / 900 s $\approx 1.3 \times 10^{12}$ cm$^{-2}$ s$^{-1}$. Considering the uncertainties in the number of sites on the Al substrate and in the adsorption probability, we approximate the H$_2$ flux to be $1$–$2 \times 10^{12}$ cm$^{-2}$ s$^{-1}$.

Using this flux of the H$_2$ beam, similar measurements were performed with a DLC sample to estimate its surface adsorption site density. At 8 K, the sum of H$_2$ signals ($J'' = 0$ and 1 levels) started to saturate after a 1500 s deposition, as shown in Figure 3b; therefore, we estimate the surface adsorption site density of the DLC sample as $1$–$2 \times 10^{12}$ cm$^{-2}$ s$^{-1}$ × 1500 s = $1.5$–$3 \times 10^{15}$ sites cm$^{-2}$. This value is close to the surface atom density of diamond, $\sim 2 \times 10^{15}$ atoms cm$^{-2}$ (Roberts & McKee 1978), indicating that the surface of our DLC sample is quite smooth.

### 3.2. TPD experiments

Figure 4 shows the REMPI-TPD spectra of H$_2$ on a DLC sample. In these experiments, the dose of H$_2$ was set to $1 \times 10^{14}$ cm$^{-2}$, which is equivalent to a relative coverage ($\theta$) of 0.067. The TPD spectra of $J'' = 0$ (*para*-H$_2$) and $J'' = 1$ (*ortho*-H$_2$) species were measured in separate experiments. The signal from $J'' = 1$ is approximately twice as strong as that from $J'' = 0$. This ratio, the ortho-to-para ratio (OPR), represents the relative population of the $J'' = 1$ species to the $J'' = 0$ species just before desorption. Because the OPR of the incident $H_2$ beam at 100 K, and consequently, that after deposition are considered to be 3, the OPR of 2 indicates that an ortho-to-para conversion occurs to some extent during the measurement period (~400 s).

For the relative coverage $\theta = 0.067$, the TPD signal starts to appear at approximately 17 K and extends to 28 K. The temperatures of the peak maxima are slightly different between the spectra of $J'' = 0$ and $J'' = 1$ species. For $\theta = 0.067$, the $J'' = 0$ spectrum shows a peak at 22.0 K, while the $J'' = 1$ spectrum shows a peak at 22.5 K. The difference in the temperatures of the peak maxima is approximately 0.5 K for TPD spectra measured for $\theta = 0.03$–$0.67$. This difference indicates that the $J'' = 1$ species is more strongly bound to the DLC sample than the $J'' = 0$ species is. We will discuss this preference in Sec. 4.2.

A series of REMPI-TPD spectra measured for various $H_2$-doses of $1$–$8 \times 10^{14}$ cm$^{-2}$, which correspond to relative coverages of $\theta = 0.067$–$0.53$, are presented in Figure 5. In these spectra, only the sum of $J'' = 0$ and 1 signals are shown. The temperature of the peak maximum for the lowest dose is at approximately 22 K. At higher doses, the temperature decreases gradually and approaches 20 K for a $H_2$ coverage of $\theta = 0.53$, whereas the profiles at the higher temperature side agree. These characteristics indicate a distribution of binding energies (desorption activation energies) over the sample surface, and $H_2$ molecules are mobile enough to find a more preferential site for adsorption prior to TPD. Further analyses of the TPD spectra will be presented in Sec. 4.1.

We measured TPD spectra after H-atom deposition (data not shown). In these experiments, the flow-rate of $H_2$ entering the microwave discharge region was kept the same as in the $H_2$-deposition experiments. For smaller doses ($\leq 2 \times 10^{14}$ cm$^{-2}$, $H_2$-equivalent), the TPD spectra obtained for H-atom deposition agree with those obtained for $H_2$-deposition within the experimental error, indicating that the adsorption coefficients are similar between H-atom and $H_2$ at 8 K, spontaneous atomic desorption is minor, and the chemisorption of H-atoms is negligible. However, from these measurements, one cannot distinguish between the following possibilities: (i) H-atoms recombine to form $H_2$ at temperatures sufficiently lower than desorption and remain on the surface, or (ii) H-atoms become mobile at elevated temperatures and the recombination reaction leads to the desorption of $H_2$. We performed another type of measurement to further understand the $H_2$ formation processes (see next section).

### 3.3. Photostimulated desorption (PSD) experiments

The number density of $H_2$ molecules on a sample surface during the deposition of $H_2$ or H-atom was monitored by the PSD-REMPI method; see Figure 1 for a schematic representation of the PSD-REMPI experiments. The temperature of the DLC sample was varied between 8 and 22 K, and experiments were also performed for $D_2$ and D-atom deposition. The experimental results for $H_2$ and H-atom deposition at 8 K are presented in Figure 6. In the beginning of $H_2$ deposition, the $J'' = 1$ (*ortho*-$H_2$) signal is stronger than the $J'' = 0$ (*para*-$H_2$) signal, reflecting an OPR of 3 for the impinging molecular $H_2$ beam. After 400 s, the signal intensity for $J'' = 0$ becomes larger than that for $J'' = 1$. These variations are presumably due to the ortho-to-para conversion of $H_2$ on the DLC surface. The sum of $J'' = 0$ and 1 signals increases almost linearly for $H_2$ fluence up to $5 \times 10^{14}$ cm$^{-2}$ (corresponding to ~400 s deposition). In the H-atom deposition experiment shown in Figure 6b, the variation in $J'' = 0$ and 1 signals is quite similar to those observed in the $H_2$ deposition experiment. These results suggest that a fraction of H-atoms readily recombine to form $H_2$ with an OPR of 3, to be discussed further in Sec. 4.3.

Figure 7 presents the results obtained for surface temperatures of 8, 18, 20, and 22 K, where only the sum of $J'' = 0$ and 1 signals is shown. At 8 K, the increase in the signal is faster for $H_2$ deposition, and the ratio of signals is approximately 1.5. This ratio becomes smaller as a function of sample temperature up to 20 K; the value is 1.25 at 18 K (Figure 7b), and it becomes almost unity at 20 K (Figure 7c). This tendency might reflect an efficient $H_2$ formation at higher temperatures. However, this tendency is not retained at 22 K, at which the ratio is 1.5, most probably due to the difference in adsorption coefficients; i.e., at this temperature the adsorption coefficient of H-atoms might be smaller than that of $H_2$, whereas the adsorption coefficients of H and $H_2$ seem to be similar at 20 K. **The enhancement of $H_2$ formation yield at higher temperatures indicates that $H_2$ formation occurs through the diffusion of physisorbed H-atoms rather than an Eley–Rideal type reaction between incoming H-atom and physisorbed H-atom.**

In $D_2$ and D-atom deposition experiments at 8 K, as shown in Figure 7e, a linear increase in the $D_2$ signal is noted to be similar to the $H_2$ and H-atom deposition experiments, but the signal intensity ratio is larger (2.0). Since no significant difference in adsorption coefficients between H-atom and D-atom is expected, this difference should be related to the diffusion rates of H- and D-atoms, in other words the rates of the recombination reactions. The signal intensity ratio remains strong even at high temperatures, i.e., 2.0 for 18–22 K, further indicating a large kinetic isotope effect in the diffusion of H- and D-atoms. The $H_2$ formation and H-atom diffusion mechanism will be discussed in Sec. 4.3.

## 4. Discussion

### 4.1. Analysis of TPD spectra

Thermal desorption is usually described by an Arrhenius-type expression that is often called the Polanyi–Wigner equation and is given as (King 1975)

$$r(\theta) = -\frac{d\theta}{dt} = \upsilon(\theta)\, \theta^n \exp[-E(\theta)/RT], \tag{1}$$

where $r$ is the rate of desorption, $\theta$ the adsorbate coverage, $t$ the time, $\upsilon$ the preexponential factor of desorption, $n$ the order of desorption, $E$ the activation energy of desorption, $R$ the gas constant, and $T$ the temperature. In the TPD experiment, temperature ($T$) and time ($t$) are related by $dT/dt = \beta$, in which $\beta$ is the heating ramp rate. Several procedures have been proposed to derive $\upsilon$ and $E$ from TPD spectra (de Jong & Niemantsverdriet 1990; Schwarz 1983). One of the popular methods is the Redhead's peak maximum method (Redhead 1962), which is applicable to coverage-independent desorption parameters and first-order kinetics. Although it is easily applicable, one needs to choose a value of $v$, and it is not applicable to systems that exhibit coverage-dependent desorption parameters.

Because the $H_2$ desorption activation energy from a DLC sample is apparently coverage-dependent, we chose to use the so-called complete analysis method (de Jong & Niemantsverdriet 1990; King 1975) to derive coverage-dependent activation energies. The procedure of this method is as follows: (i) TPD spectra are integrated from the higher temperature side, and the initial coverages are determined; (ii) for each integrated curve, find a temperature ($T$) at which a certain coverage ($\theta'$) is reached; (iii) for this coverage (i.e., $\theta'$) one obtains a set of ($r$, $T$) from the TPD spectra; (iv) make an Arrhenius plot of all $r$ against $1/T$, which yields an activation energy ($E(\theta')$) from the slope of the plot; and (v) repeat (i)–(iv) for different coverages to obtain the coverage-dependent activation energy ($E(\theta)$). The preexponential factor ($v$) is, in principle, obtained from the intercept $n \ln \theta' + \ln v(\theta')$ when the order of the desorption is known. However, because the intercept obtained from REMPI-TPD spectra contains a proportionality factor ($A$), which relates the desorption rate and REMPI-signal intensity ($S(\theta)$) as $S(\theta) = A \times r(\theta)$ (that is, the intercept becomes $n \ln \theta' + \ln v(\theta') + \ln A$), and because the complete analysis method does not allow to confidently constrain the preexponential factor to within less than a few orders of magnitude (Amiaud et al. 2006), we did not attempt to accurately determine it.

Figure 8 shows the activation energies obtained from the complete analysis for the three sets of TPD spectra. The activation energy lays at approximately 30 meV for a range of relative coverage up to 0.2, and it becomes smaller as the relative coverage increases. Overall, the activation energy distribution is quite narrow in the investigated range of relative coverage, in contrast to the variation in activation energies (40–60 or 45–65 meV) demonstrated for $D_2$ desorption from the ASW surface (Amiaud et al. 2006; Hornekær et al. 2005).

We found that the product of the preexponential factor and proportionality factor, ν × A, is nearly constant for the range of relative coverage (0.02–0.4). If we assume that this product is coverage-independent, we can convert eq. (1) for $E(\theta)$ to obtain

$$E(\theta) = RT \ln\left[\frac{\nu\theta}{r(\theta)}\right]. \tag{2}$$

By using the relations, $S(\theta) = A \times r(\theta)$ and $dT/dt = \beta$, eq. (2) is written as

$$E(\theta) = RT \ln\left[\frac{A\nu\theta}{\beta S(\theta)}\right]. \tag{3}$$

Using this equation, $E(\theta)$ was obtained from the TPD spectrum with an initial coverage of 0.53 and is represented in Figure 8 by a solid line. This curve agrees well with activation energies obtained from the complete analysis. It indicates that at coverages below 0.04, the activation energy rises as high as 40 meV, although the preexponential factor may be different in this region. However, assuming a typical preexponential factor for the first order desorption to be in a range $10^{12}$ to $10^{13}$ s$^{-1}$, the errors originating from the uncertainty in the preexponential factor are not significant (< 4 meV).

To our knowledge, only one experimental study on molecular hydrogen desorption from **an amorphous** carbonaceous surface has been reported (Pirronello et al. 1999). Pirronello et al. reported TPD spectra of HD from a graphitic amorphous carbon sample, where HD was produced from H- and D-atoms, and determined the overall activation energy for HD formation to be ~45 meV; in their analysis, activation energies for recombination and desorption were not separated, and a coverage-independent preexponential factor and activation energy were assumed. Later, Katz et al. analyzed these TPD data and obtained activation energies of 44.0 and 46.7 meV for the atomic diffusion and molecular desorption, respectively (Katz et al. 1999). Although Katz et al. did not include coverage-dependent terms in their model, the TPD spectra of Pirronello et al. were reasonably reproduced. After all, the activation energy for molecular ($H_2$ or HD) desorption is larger for a graphitic amorphous carbon than our DLC sample. Although one cannot directly compare TPD spectra measured with a different ramp rate ($\beta$), the similarity between their and our TPD spectra implies that the distribution of activation energies is also narrow for the graphitic amorphous carbon sample. A relatively wide distribution of desorption activation energies has been reported for $D_2$ desorption from ASW. Hornekær et al. (Hornekær et al. 2005) reported activation energies of 40–60 meV, while Amiaud et al. (Amiaud et al. 2006) reported 45–65 meV; thus, the energies reported by these groups agree well with each other. Porous rough surfaces tend to have a large distribution of activation energies, and in turn, the narrow distribution found in this work indicates the less porous morphology of our DLC sample. Moreover, this conclusion is consistent with the adsorption site density for our DLC sample (1.5–3 × $10^{15}$ sites $cm^{-2}$) being significantly smaller than that of a porous ASW sample (5–10 × $10^{15}$ sites $cm^{-2}$) (Al-Halabi & Van Dishoeck 2007; Hidaka et al. 2008).

**The obtained activation energy for $H_2$ desorption from the DLC surface, ~30 meV, is much smaller than that reported for $H_2$ desorption from a graphite surface, 41 meV (Mattera et al. 1980). Considering similarities in the van der Waals interaction between helium-graphite and helium-diamond (111) (Vidali et al. 1991), the activation energy ~40 meV may be expected for $H_2$ desorption from the diamond (111) surface. The reduction of van der Waals force between helium and diamond (111) surface upon hydrogenation has been reported (Su & Lin 1998; Vidali et al. 1983). However, it is not the case for our DLC sample as we demonstrated that the fraction of hydrogenated site is quite small. Therefore, the smaller activation energy for $H_2$ desorption from the DLC surface is thought to originate from amorphous nature of our DLC sample. The determination of van der Waals interaction between helium and DLC surface is desirable to clarify the origin of differences.**

**4.2. Rotational-state dependence of binding energy**

The TPD spectrum of the $J'' = 1$ species at a certain coverage is slightly shifted towards higher temperature compared to that of the $J'' = 0$ species, with a temperature difference of approximately 0.5 K between the peak maxima, indicating that the $J'' = 1$ species is more strongly bound to the surface. We could not obtain the rotational-state-dependent activation energies from the complete analysis because the coverage of each species is not well defined due to the partial ortho-to-para conversion. Instead, TPD spectra were simulated according to eq. (1) with coverage-independent $E$ and $\nu$, and the observed difference in peak temperature is reproduced with an activation energy difference of *ca.* 1 meV.

Such an activation energy difference between *para-* and *ortho-*$H_2$ desorption has been reported in the literature (Fukutani & Sugimoto 2013). By using the REMPI-TPD method, Fukutani and co-workers determined the rotational-state-dependent activation energies of $H_2$ on activated $Al_2O_3$ (Magome, Fukutani, & Okano 1999) and on Ag(111) (Sugimoto & Fukutani 2014) surfaces, and Amiaud et al. reported that of $D_2$ on ASW (Amiaud et al. 2008). The difference has been attributed to an anisotropy in the potential energy surface, which leads to rotational sublevel splitting in *ortho-*$H_2$ (Fukutani & Sugimoto 2013). The estimated difference for $H_2$ on DLC, ~1 meV, is similar to those reported for $D_2$ on ASW (1.4 ± 0.3 meV) (Amiaud et al. 2008) and for $H_2$ on Ag (2 meV) (Sugimoto & Fukutani 2014).

### 4.3. H-atom diffusion and $H_2$ formation

For the ASW and solid CO surfaces, H-atom diffusion has been studied by measuring time variations in the surface H-atom density by the PSD-REMPI method (Hama et al. 2012; Kimura et al. 2018; Watanabe et al. 2010). However, we found that H-atoms on a DLC surface could not be detected by this method, probably because of a very low PSD efficiency. Instead, we performed PSD-REMPI measurements of surface $H_2$ to extract information on H-atom diffusion ($H_2$ recombination).

In the PSD-REMPI experiments shown in Figures 6 and 7, we demonstrated that the sum of $H_2$ $J'' = 0$ and 1 signals upon H-atom deposition at 8 K is weaker than that upon $H_2$ deposition with a ratio of 1.5. When similar measurements were performed with an Al substrate at 8 K in the previous experiment (Hama et al. 2012), a more significant difference, a ratio of up to 2.0, was seen and this observation has been explained as follows: the $H_2$ signal observed for H-atom deposition originates from undissociated $H_2$ contained in the H-atom beam and recombined $H_2$ that remained on the surface. The 1.5 ratio observed for the DLC sample at 8 K cannot only be due to undissociated $H_2$, considering a dissociated fraction of up to 70% and a ratio smaller than that observed for the Al substrate; therefore, we think that a fraction of adsorbed H-atoms remains unreacted, while the recombined and undissociated $H_2$ stays on the surface. This hypothesis is supported by the REMPI-TPD experiments, in which the TPD spectra observed for $H_2$ deposition are nearly identical to that for H-atom deposition (equivalent dose). Moreover, the agreement between TPD spectra, especially in shape, indicates that the orders of desorption are the same (first order) for $H_2$ and H-atom depositions; therefore, $H_2$ recombination occurs at a temperature below desorption. The time variation patterns for the $J'' = 0$ and 1 signals were very similar between $H_2$ and H-atom deposition PSD-REMPI experiments, as shown in Figure 6. Because the time variation in the signal during the $H_2$ deposition results from the ortho-to-para conversion occurring on the DLC surface, during the deposition of the molecular $H_2$ beam at an OPR of 3, the similarity in the pattern suggests that the nascent OPR of the recombined $H_2$ follows the statistical value of 3, as found in the ASW case (**Gavilan et al. 2012**; Watanabe et al. 2010).

Above, we suggested that $H_2$ recombination became efficient at elevated temperatures, and in the following, the temperature dependence of $H_2$ recombination will be discussed quantitatively. In the PSD-REMPI experiments, the $H_2$ signal intensities are proportional to the surface number density of $H_2$. Hereafter, the $H_2$ number densities during $H_2$ deposition and H-atom deposition are represented with $n'(H_2)$ and $n(H_2)$, respectively. We discuss $H_2$ recombination using the signal intensity ratio (SIR), which is derived from the experiments and is represented by $n'(H_2)$ and $n(H_2)$ as

$$\text{SIR} = \frac{n'(H_2)}{n(H_2)}. \tag{4}$$

From the SIR, we can derive the ratio of the surface number densities of $H_2$ and H-atoms during H-atom deposition, $n(H_2)$ and $n(H)$, and using this ratio, we would like to derive the $H_2$ recombination yield of H-atoms deposited, which is defined as [the number of recombined H-atoms] / [total number of H-atoms deposited on the surface]. Because the same gas flow rate ($H_2$ equivalent) was used in the $H_2$ and H-atom deposition experiments and because the adsorption coefficients of $H_2$ and H-atom are assumed to be similar, the surface number densities of $H_2$ during $H_2$ deposition, $n'(H_2)$, can be expressed using those of $H_2$ and H-atom during the H-atom deposition, $n(H_2)$ and $n(H)$, as $n'(H_2) = n(H_2) + n(H)/2$. By substituting this relation into eq. (4), we derive the following relation

$$\text{SIR} = \frac{n(H_2) + n(H)/2}{n(H_2)} = 1 + \frac{1}{2}\frac{n(H)}{n(H_2)}. \tag{5}$$

Equation (5) reveals that the SIR value is related to the ratio between $n(H_2)$ and $n(H)$. Assuming that the dissociated fraction in the atomic hydrogen beam is 70% and that no recombination occurs at the sample surface, the ratio of the surface number densities of $H_2$ and H-atom becomes $n(H_2):n(H) = 30:140$, leading to an SIR of 3.33. When the recombination yield is $x$% and the recombined $H_2$ stays on the surface, the ratio of the surface number densities is $n(H_2):n(H) = (30 + 0.7x):(140 - 1.4x)$; using this ratio, the SIR becomes $100 / (30 + 0.7x)$. Thus, the SIR of 1.5 at 8 K indicates that ~50% of H-atoms recombined to form $H_2$.

The SIR upon $H_2$ and H-atom deposition became smaller when the sample temperature increased (see Figure 7): 1.5 at 8 K, 1.25 at 18 K, and 1.0 at 20 K, corresponding to 50%, 70%, and 100% recombination yields. These variations **can be explained by thermal activation of H-atom diffusion not by Eley–Rediel type reactions and indicate** that the produced $H_2$ stays on the surface without desorption upon recombination. It is of note that the ratio is nearly independent of the deposition time, in other words, the coverage. The 100% recombination yield at 20 K indicates that H-atoms can encounter reaction partners after a rapid diffusion even at low coverage. **The retention of recombined $H_2$ on the DLC surface was observed in our experiments; however, we cannot exclude the possibility that a small fraction of recombined $H_2$ desorbs right away. The $H_2$ (or HD) desorbing from a graphite surface has been detected by Price and co-workers (Creighan, Perry, & Price 2006; Islam, Latimer, & Price 2007). In their experiments, only $H_2$ desorbing from the surface was measured by the REMPI method; i.e., they are not able to quantify the fraction of desorbed $H_2$ relative to the $H_2$ remaining on the surface. Nevertheless, to explain the $H_2$ retention on the DLC surface, the energy dissipation processes should be considered rather than the trapping of $H_2$ in a morphologically complex surface (Hornekær et al. 2003; Roser et al. 2002). Further investigations, especially theoretical ones, are required to explain these observations.**

To discuss $H_2$ recombination under astrophysically relevant conditions, the residence time of H-atoms on a dust grain surface is rather important since the accretion rate of H-atoms is very slow—it is approximately one H-atom per day for a dust size of 0.1 μm with an H-atom density of 1 $cm^{-3}$. If we ignore the loss of H-atoms by chemical reactions, the residence time, τ, is determined by the desorption activation energy ($E_{des}$) of the H-atom and the temperature ($T$) as

$$\tau = \left[\nu_{des}\exp(-\frac{E_{des}}{k_B T})\right]^{-1}, \tag{6}$$

where $\nu_{des}$ is the frequency factor for desorption, with a typical value of $10^{13}$ $s^{-1}$. According to eq. (6), $E_{des}$ = 71 meV is required to achieve the residence time of one day at 20 K. Such a high activation energy (i.e., binding energy) might not be possible for H-atom physisorption systems. Therefore, we think that $H_2$ recombination at 20 K on a DLC surface is unlikely to occur under astronomical conditions, although a determination of the activation energy is still necessary for confirmation. For a graphitic amorphous carbon surface, $E_{des}$ is reported to be 56.7 meV (Katz et al. 1999). This value was derived from simulating TPD spectra by using three types of free parameters: diffusion rates, desorption rates, and the fraction of $H_2$ molecules that remains on the surface upon formation. The $E_{des}$ for a DLC surface is expected to be lower than 56.7 meV considering that the $H_2$ desorption activation energy (~30 meV) determined for a DLC surface is smaller than the 46.7 meV estimated for a graphitic amorphous carbon surface (Katz et al. 1999). Actually, even for $E_{des}$ = 56.7 meV, a temperature as low as 15 K is required for the $H_2$ formation to occur under astrophysical conditions. Therefore, chemisorbed H-atoms might play a role in $H_2$ formation at relatively high temperatures (> 20 K) (Cazaux & Tielens 2004).

One of the interesting findings is the significant isotope effect in $H_2$ and $D_2$ formation. The ratio of the sum of $D_2$ $J'' = 0$ and 1 signals upon $D_2$ deposition to that upon D-atom deposition (i.e., SIR) is almost constant at 2.0 for the temperature range 8–22 K, as shown in Figures 7e–g. These results imply that the fraction of D-atoms that remain unreacted on the surface is constant (~70%) for this temperature range, in contrast to the efficient diffusion of H-atoms at a higher temperature of ~20 K. Such a large kinetic isotope effect on diffusion may indicate quantum-tunneling because the tunneling probability strongly depends on the tunneling particle mass. Recently Kuwahata et al. reported evidence of the quantum-tunneling diffusion of H-atoms on polycrystalline water ice (Kuwahata et al. 2015), where a significant isotope effect is found in contrast to the small isotope effect in the diffusion of H- and D-atoms on ASW (Hama et al. 2012). According to theoretical studies, H-atom diffusion on ASW is suppressed due to its nonperiodic potential (Smoluchowski 1979, 1981, 1983). In other words, quantum-tunneling diffusion is thought to dominate within a single crystal, whose potential is relatively shallow and regular. Further evidence for quantum-tunneling diffusion can be provided by performing the experiments and analyses developed by Kuwahata et al. (Kuwahata et al. 2015). In their method, the ratio of the surface number densities of H-atoms and D-atoms on the surface during atomic deposition, $n(D) / n(H)$, is a measure of the difference in diffusion rate constants. An inability to detect H-atoms on the DLC surface by the PSD-REMPI method does not allow us to present direct evidence of quantum-tunneling diffusion.

In the following, we will suggest that the observed isotope effect cannot be explained by only considering thermal hopping diffusion. When the surface diffusion of H(D)-atoms is limited to thermal hopping, the steady-state $n(D) / n(H)$ ratio at a certain H(D)-atom flux is expressed with the rate constants of the recombination reaction as

$$\frac{n(D)}{n(H)} = \sqrt{\frac{k_{H+H}}{k_{D+D}}}, \tag{7}$$

where $k_{H+H}$ and $k_{D+D}$ are recombination rate constants for H- and D-atoms, respectively (see note[1] for the derivation of eq. (7)). Because the recombination reaction is a radical-radical barrier-less reaction, its rate constant ($k_{H+H}$) is generally determined by H-atom diffusion and is expressed as

$$k_{H+H} = s\nu_H \exp(-\frac{E_{diff,H}}{k_B T}), \tag{8}$$

where $s$ is the unit area of the surface site, $\nu$ is the frequency factor, and $E_{diff,H}$ is the diffusion activation energy of H-atoms. When the mass of diffusing particle is significantly smaller than that of the surface species, the frequency factor ($\nu$) is generally proportional to the inverse of the square root of the mass (Glyde 1969; Prigogine & Bak 1959). Therefore, the ratio of rates for H-atoms and D-atoms is written as

$$\frac{k_{H+H}}{k_{D+D}} = \sqrt{2}\exp(-\frac{\Delta E_{diff,D-H}}{k_B T}), \tag{9}$$

where $\Delta E_{diff,D-H} = E_{diff,D} - E_{diff,H}$. In practice, $\Delta E_{diff,D-H}$ originates from the difference in zero-point energy for thermal hopping. By using the $\Delta E_{diff,D-H} = 1$ meV determined for the ASW surface (Hama et al. 2012), eq. (9) gives $k_{H+H} / k_{D+D} \approx 2.5$ at 20 K. Consequently, using eq. (7), $n(D) / n(H) = 1.6$ is obtained for $T = 20$ K.

In the present experiments, we cannot directly determine the $n(D) / n(H)$ ratio, but it can be estimated from separate experiments using the same H- and D-atom flux. First, we calculate the surface number density ratio for $D_2$ and D-atoms from the SIR of 2.0 measured for $D_2$ and D-atom depositions at 20 K (Figure 7g). According to eq. (5), the ratio between the surface number densities of $D_2$ and D-atoms is derived as $n(D_2):n(D) = 1:2$ at 20 K. Next, we estimate the signal intensity ratio expected for $H_2$ and H-atom depositions at 20 K. By using the $n(D) / n(H) = 1.6$ estimated above, the ratio of surface $H_2$ and H-atoms during H-atom deposition, at the same flux as that for the D-atom deposition experiment, is calculated to be $n(H_2):n(H) = 1:0.9$ (see note[2] for the derivation of this ratio); consequently, the signal intensity ratio for $H_2$ and H-atom depositions at 20 K should become ~1.5, according to eq. (5). This value is inconsistent with the experimentally determined ratio of ~1.0 (Figure 7c)). The inconsistency should originate from the assumption of thermal hopping diffusion; i.e., H-atom diffusion only with the thermal hopping mechanism cannot explain the absence of H-atoms and the significant fraction of unreacted D-atoms at 20 K. Therefore, the quantum-tunneling diffusion mechanism should be taken into account for H-atom diffusion on a DLC surface. Because an enhancement in $H_2$ recombination was observed at elevated temperatures, we suggest that H-atom diffusion on a DLC surface occurs through a thermally assisted tunneling mechanism with partial contribution from thermal hopping. Further investigations are required to discuss the percent contributions of tunneling and thermal hopping diffusions.

**Notes**

1. The time variation in the surface number density of H-atoms is expressed by a rate equation as

   $dn(H)/dt = -k_{H+H}n(H)^2 - k_{des}n(H) + p_sF$, where $k_{des}$ is the rate constant for monoatomic desorption, $p_s$ the adsorption coefficient, and $F$ the flux of H(D)-atom. Since monoatomic desorption is negligible, according to TPD experiments, at steady state with $dn(H) / dt = 0$, the rate equation becomes $n(H) = (p_s F / k_{H+H})^{1/2}$. Thus, when the flux of H-atoms is the same as that of D-atoms, eq. (4) is derived.

2. More specifically, this ratio was derived by the following calculations: first, we assume that the number of impinging particles is 100 ($H_2$ or $D_2$ equivalent). Then, we derive $n(D_2) = 50$ and $n(D) = 100$ to satisfy the relation between the observed intensity ratio and number densities: $2.0 = [n(D_2) + n(D)/2] / n(D_2)$, eq. (4). By using $n(D) / n(H) = 1.6$, the number of H-atoms is calculated to be $n(H) = 62.5$. Since $n(H_2) + n(H)/2 = 100$ is assumed, $n(H_2)$ becomes 68.8 and we derive $n(H_2):n(H) = 68.8:62.5 \approx 1:0.9$.

**Acknowledgements**

This work was supported by a JSPS Grant-in-Aid for Specially Promoted Research (JP17H06087) and partly by a JSPS KAKENHI (JP18K03717).

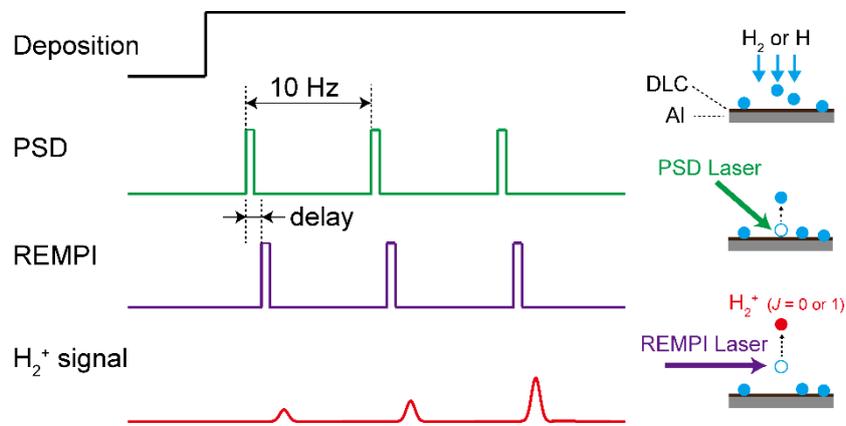

Figure 1. Schematic illustration of the PSD-REMPI experiments. The timing chart is presented on the left side. The deposition of $H_2$ or H-atoms, desorption of $H_2$ by the PSD laser, and ionization of $H_2$ by the REMPI laser are illustrated on the right side. Because $H_2$ or H-atoms are continuously deposited during a measurement, the $H_2^+$ signal intensity increases as a function of time.

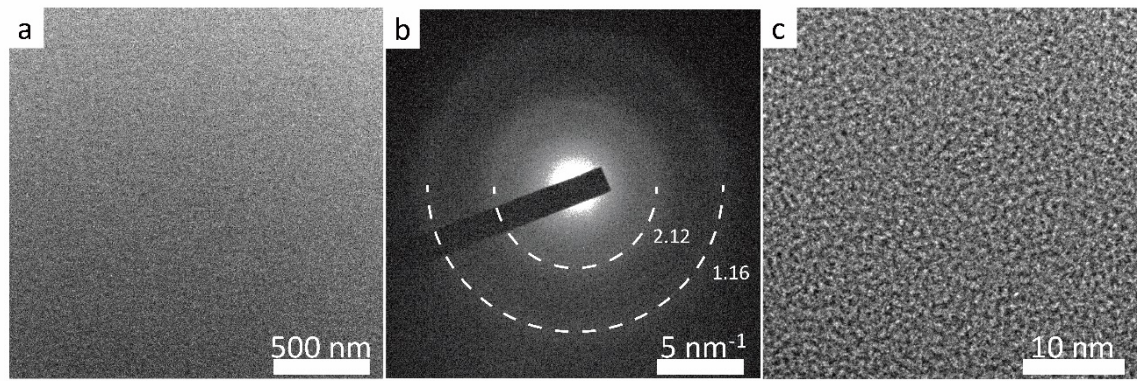

Figure 2. Transmission electron micrographs of a free standing DLC thin film on a standard copper TEM grid. a: A typical bright-field image. b: The corresponding electron diffraction pattern. c: A high-resolution image. The dotted semicircles in b indicate the positions of the Debye rings corresponding to 2.12 and 1.16 Å.

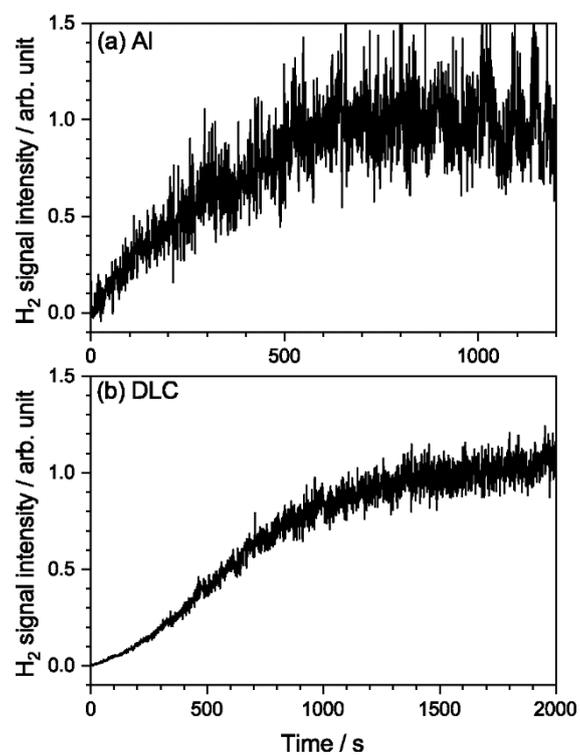

Figure 3. Evolution of H$_2$ (the sum of $J'' = 0$ & 1) signals measured with the PSD-REMPI method for the (a) Al substrate and (b) DLC sample.

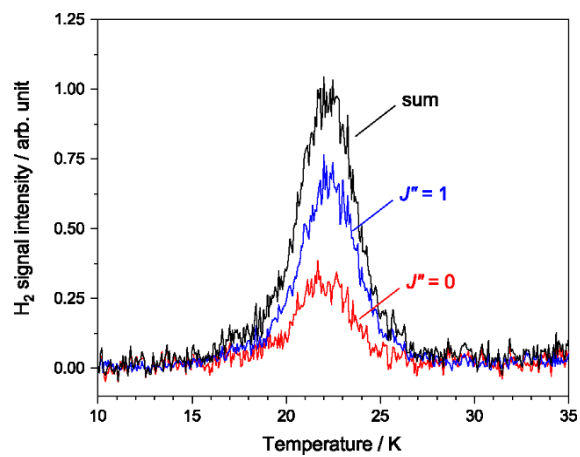

Figure 4. REMPI-TPD spectra measured for $H_2$ desorbing from a DLC surface with an initial relative coverage $\theta = 0.067$. $H_2$ molecules were deposited at 8 K for 100 s with a flux of $1–2 \times 10^{12}$ cm$^{-2}$ s$^{-1}$. TPD runs were started 50 s after finishing the deposition. The heating rate was 4 K min$^{-1}$. The REMPI-TPD spectra of $J'' = 0$ species, $J'' = 1$ species, and their sum are shown by red, blue, and black lines, respectively.

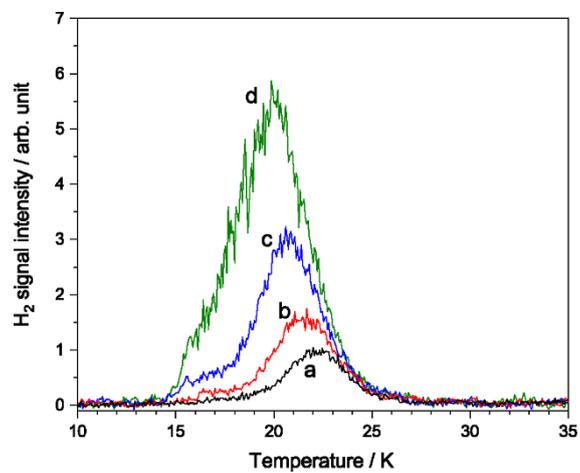

Figure 5. Initial coverage dependence of the REMPI-TPD spectra. The sum of H$_2$ ($J'' = 0$ & 1) signals are shown for initial relative coverages ($\theta$) of (a) 0.067, (b) 0.13, (c) 0.27, and (d) 0.53. H$_2$ molecules were deposited at 8 K for a certain period with a flux of 1–2 × 10$^{12}$ cm$^{-2}$ s$^{-1}$. TPD runs were started 50 s after finishing the deposition. The heating rate was 4 K min$^{-1}$.

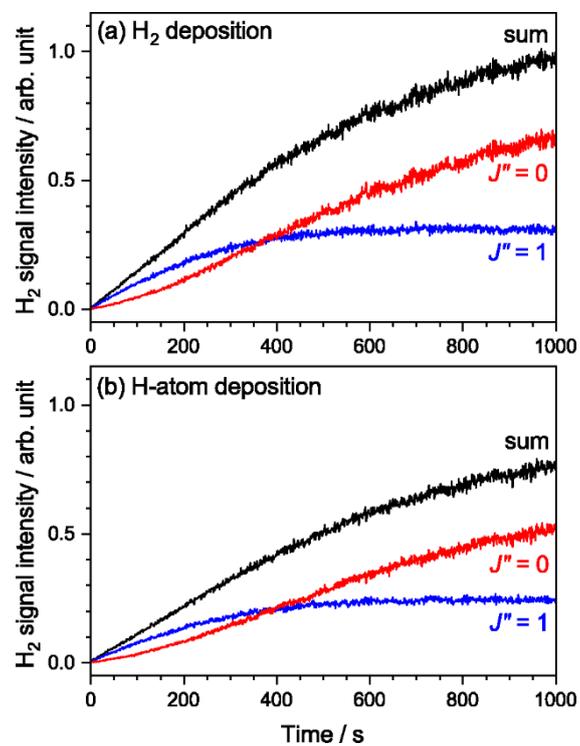

Figure 6. Time variation of PSD-REMPI signals measured during (a) $H_2$ and (b) H-atom deposition. The $J'' = 0$ and 1 signals are shown by red and blue lines, respectively, and the sum of $J'' = 0$ and 1 signals is shown by the black line. The sample temperature was 8 K.

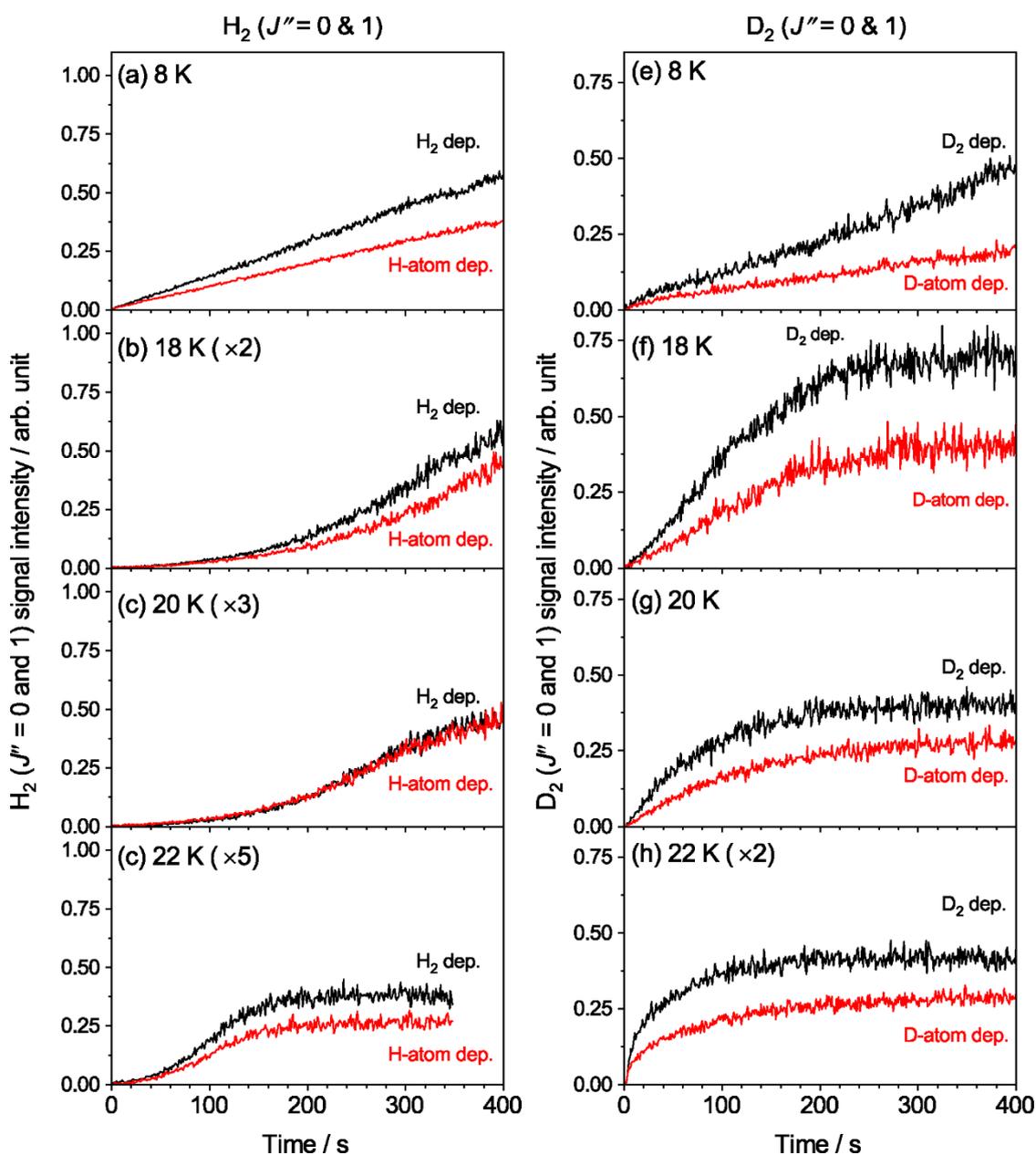

Figure 7. Time variation of PSD-REMPI signals measured during (left column) $H_2$ or H-atom deposition and (right column) $D_2$ or D-atom deposition. From top to the bottom, the sample temperatures were 8, 18, 20, and 22 K, respectively. The signals following molecular deposition are shown in black lines, and those following atomic deposition are in red lines. The sum of $J'' = 0$ and 1 are presented, and each trace was obtained as an average of three measurements. For scaled traces, the multiplication factors are shown in parentheses.

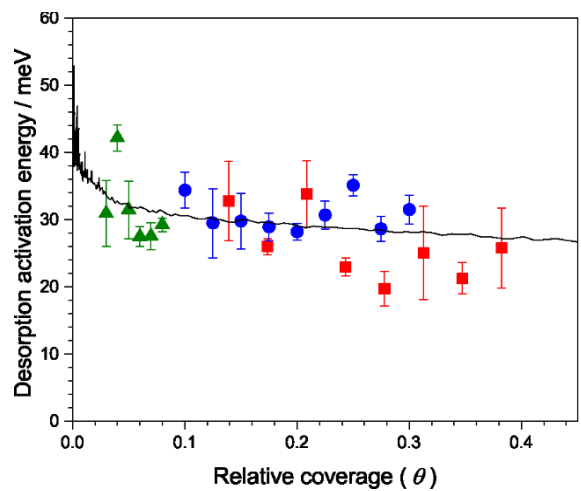

Figure 8. Coverage-dependent activation energy for $H_2$ desorption determined for DLC samples. Activation energies shown by filled triangles, circles, and squares were determined by the complete analysis method for three sets of TPD spectra. Error-bars represent those in linear fitting to Arrhenius type plots. The solid line was obtained according to eq. (3) and the TPD spectrum measured for $\theta = 0.53$; see text for details.